\documentclass[runningheads]{llncs}
\usepackage{graphicx}
\usepackage{caption}
\usepackage{subcaption}
\usepackage{graphicx}
\usepackage{xcolor}
\usepackage{hyperref}
\newcommand{\jbc}[1]{{\color{black} #1}}

\begin{document}

\title{ETL for the integration of remote sensing data} 
\titlerunning{ETL for remote sensing}
%
\author{Romero Jure, P. V.\inst{1,2}\orcidID{0000-0001-5892-2137}\email{paula.romero@mi.unc.edu.ar} \and
Cabral, J. B.\inst{1,3}\orcidID{0000-0002-7351-0680}\email{jbcabral@unc.edu.ar} 
\and Masuelli, S.\inst{2}\email{smasuelli@unc.edu.ar} }
\authorrunning{Romero et al.}
%
\institute{
Gerencia de Vinculación Tecnológica, Centro Espacial Teófilo Tabanera, CONAE sede Córdoba, Argentina\\ \and
Facultad de Matemática, Astronomía, Física y Computación; Universidad Nacional de Córdoba, Córdoba, Argentina\\
\and Instituto de Astronomía Teórica y Experimental, Observatorio Astronómico de Córdoba, Córdoba, Argentina}
\maketitle              
\begin{abstract}
\jbc{Modern in-orbit satellites and other available remote sensing tools have generated a huge availability of public data waiting to be exploited in different formats hosted on different servers. In this context, ETL formalism becomes relevant for the integration and analysis of the combined information from all these sources.
Throughout this work, we present the theoretical and practical foundations to build a modular analysis infrastructure that allows the creation of ETLs to download, transform and integrate data coming from different instruments in different formats. Part of this work is already implemented in a Python library which is intended to be integrated into already available workflow management tools based on acyclic-directed graphs which also have different adapters to impact the combined data in different warehouses.}



\end{abstract}
\section{Introduction}

\jbc{The Extraction, Transformation, and Loading (ETL),  is the formalism for  extracting data from various sources, transforming it into a useful format, and loading it into a target repository, such as a data warehouse.}

\jbc{The term gained popularity throughout the industry around the 1970s rather than being formally defined in a document.}
However, previous works have settled the bases for the formalism. One of the first works is \cite{etl}, which widely describes the process and its relation with Data Warehouse. Furthermore, \cite{bestETL} defines ETL activities and provides formal foundations for its conceptual representation.

\jbc{ETL serves as a practical theoretical framework for data integration by simplifying the extraction of data from different sources, their transformation into a consistent and compatible format, and their loading into a centralized data warehouse to facilitate subsequent analysis.}

In this context, remote sensing instruments on board artificial satellites orbiting the Earth are generating huge amounts of data every day, which is considered to be a Big Data problem \cite{etl_remsens} \cite{sat_data_asBD}. The data is transmitted to Earth, stored in a data warehouse system and usually provided to the user in some scientific file format, such as a Network Common Data Form (NetCDF) \cite{rew2006netcdf}, Hierichal Data Format \cite{hdf} or GeoTIFF\cite{geotiff}. The database where the files are stored and the format depends on the agency or organization responsible for each satellite. There exist several situations where someone needs to analyze Earth Observation data by combining measurements from multiple instruments onboard different satellites. That would be an appropriate problem for ETL formalism because we would be dealing with big data stored in different sources and we need to transform it into a product and load the latter in some database. 

An event where different (satellite) sensors observe the same location roughly at the same time is called a collocation \cite{collocations}. Several works have required implementing the collocation finding procedure, for example, \cite{cumulo} generated a dataset with combined data from MODIS and the Cloud Profiling Radar (CPR) on-board CloudSat, to study cloud types. \cite{collocations} have studied collocations between the Microwave Humidity Sounder (MHS) on-board NOAA-18 and the CPR. In practice, collocating Earth Observation data usually comes with many problems due to the different sources where the data is stored and the compatibility of data formats. 

\jbc{In this context, we have decided to use the ETL formalism to integrate remote sensing data from multiple sources by designing extractors, transformers and loaders that access information from platforms provided by different missions. Although there are precedents of application \cite{etl_remsens}, in our work we opted for a modular mechanism so that different users can customize their processing and analysis pipelines.

Although there are several programs suitable for performing collocations, particularly Geographic Information System (GIS) software, we have decided to implement all this infrastructure in Python, given its popularity and ecosystem in scientific computing \cite{python}.}
Besides, even though there exist some Python libraries that implement data processing methods for Earth Observation data such as \textit{Satpy} \cite{satpy}, we have not found any that implement extraction methods and integrate them with the processing stage, most of them assume that the data is available in the local system.

%

  
\jbc{This paper is organized as follows: 
In Section \ref{sec: formalism} we present the ETL formalism and its relation to remote sensing data, then. In Section \ref{sec: design} we present and explain our design of a general ETL modular pipeline intended to combine data from instruments on board different artificial satellites in orbit around the Earth. In Section \ref{sec: design} we present an implementation of the design as a Python package.  In Section \ref{sec: results} we present the results in the former and in Section \ref{sec: conclusion}, conclusions and future work to be done.}



\section{ETL formalism}\label{sec: formalism}

\jbc{The acronym ETL (Extract, Transform, Load) emerged in the context of data warehousing around the 1970s.
And it comprises the following stages: Extracting data from the original sources, quality assuring and cleaning data, conforming the labels and measures in the data to achieve consistency across the original sources, and delivering data in a physical form \cite{etl}.}


This \jbc{stages} can be represented generally as a kind of 
diagram. For example \cite{bestETL} have developed a graphical notation useful to ``capture the semantics of the entities involved in the ETL process". We have adopted this notation to represent the design proposed. In the following paragraph, we will review some basic concepts that are needed to explain our proposed design. 

\begin{figure}
    \centering
    \includegraphics[width=12cm,height=3.5cm]{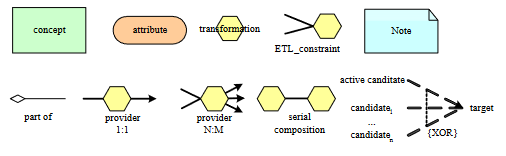}
    \caption{Graphical notation for explaining an ETL pipeline. Figure courtesy of \cite{bestETL}.}
    \label{fig:notacion}
\end{figure}

\jbc{
Figure~\ref{fig:notacion} shows all the elements that could be present in the diagram: 

The \textit{Attributes}, which are the minimum unit of information, represented with an oval shape. The \textit{Concepts}, squares, are the entities of the source databases and are defined by a name and a finite set of attributes. The hexagons represent the \textit{Transformations}, which are the parts of code that execute a task. 
Next, the \textit{ETL\_Constrains} are a finite set of attributes, on which the constraint is imposed and a single transformation that implements the enforcement of the constraint. Finally, the \textit{Notes} contains comments.

It is important to note that all notation is UML\cite{uml} based but some forms do not have the same meaning.

In this work, we will focus on transformation operations that transform the input data.
}

\section{Design }\label{sec: design}

\jbc{
Our design is based on \cite{bestETL} and is graphically represented in Figure~\ref{fig:sub1}. For simplicity we have restricted our analysis to two sources/instruments/databases, one containing the data transmitted by satellite ``A" and the other, by satellite ``B''.  Usually, each database contains different products with several levels of processing, but all of them share the format with some common attributes and metadata.
}

As stated in \cite{collocations}, to have a meaningful collocation, the pixels from both images must have a physical overlap, which means that they need to meet a spatial and a temporal criterion within some threshold of error.

First, two files that may meet the time overlap criterion are selected and downloaded.
\jbc{
For example: An Extractor retrieves an Image \textit{A} that was stored in format \textit{A} from Data source \textit{A}, so the file can be named ``ImageA.extA", where ``extA'' means the extension for file format \textit{A}. The more  common scientific file formats for Earth Observation data are NetCDF (\texttt{.nc})\cite{rew2006netcdf}, a version of HDF (\texttt{.hdf}), \texttt{.h5})\cite{hdf}, and GeoTIFF (\texttt{.tiff})\cite{geotiff}.

Each Image is for our formalism a \textit{Concept} characterised by the \textit{Attributes}: 
}
\begin{description}
    \item[Time:] The time at which each pixel of the image was acquired. It is usually provided in UTC or in some format of absolute time with a defined origin.
    \item[Geoloc:] The geolocation of each pixel of the image, the point on Earth measured by the sensor. Each pixel is characterised by its spatial coordinates in some projection related to the type of orbit of the satellite. 
    \item[Parameters:] Every pixel of the image is characterized by n parameters. A parameter is a measurement taken by the satellite instrument or some quantity derived from it.
\end{description}

A \textit{Transformer} $c$ transforms the files into a common format so it will be easier to work with them and perform the collocations later. Formally

\jbc{
$$
    c : extA, extB \rightarrow extC 
$$

Then, the \textit{Transformer} $f$ gets the location of every pixel from image A as input and converts its coordinates from projection A to Projection B. 

$$
    f : coordA_{projA} \rightarrow coordA_{projB}
$$
}
Next, pixels that meet the spatial overlap criterion within some threshold of error are selected and can be collocated. \textit{Transformer} $t$ takes care of that task, taking the coordinates of geolocation, both in some projection, as input and retrieving a new product as output. 

$$
    t: Parameters A, Parameters B \rightarrow PixelA\&B
$$
\jbc{
The output is a \textit{Concept} called ``Pixels A\&B", which format is the common format, and contains the Parameters from A and the Parameters from B, that were attributes of the pixels A and B and have not been altered or transformed. The entire process can be represented as

$$
    (c \circ f \circ t): ImgA, ImgB \rightarrow PixelA\&B
$$

Finally, a \textit{Loader} loads the final product into a Database.
}

\section{Results: An implementation}\label{sec: results}

\jbc{
We have applied the discussed design in a Python Package that, a priori, aims to be used to collocate \cite{collocations} data from a radiometer aboard a geostationary satellite with data from a radar aboard a polar-orbiting satellite.}

\jbc{
For the source ``A" of data we chose the ABI (Advanced Baseline Imager) on board geostationary GOES-16, with a central longitude of -76°, which allows it to take images of the whole American continent, with a temporal frequency ranging from 5 to 15 minutes \cite{goesoficial}. 
The ABI is a multispectral radiometer that sense the Earth in 16 bands ranging from the visible to the NIR part of the electromagnetic spectrum \cite{tj}. 
An image of the whole continent is square and has a spatial resolution of 0.5 km to 2 km and a side size of 5424 to 16272 pixels, depending on the band. See table \ref{tab: goes bands} for details about this.} 
Every pixel of an image is geolocated and each file name contains information about the acquisition time of the measurements. 
The ABI data is stored in \textit{NetCDF} format hosted in  Amazon Web Server (AWS) \cite{rew2006netcdf}.

\begin{center}
\begin{tabular}{ r| r| r } \label{tab: goes bands}
 Bands & Resolution (Km) & Image size (pixels) \\
 \hline
 2 & 0.5 & 16272 \\ 
 1, 3, 5 & 1 & 10848 \\ 
 4, 6-16 & 2 & 5424 \\ 
\end{tabular}
\end{center}

\jbc{
As source ``B'' we choose the CPR (Cloud Profiler Radar) on board the polar satellite CloudSat. 
Its orbit is polar sun-synchronous, with a temporal frequency of 16 days.
The CPR was designed to generate vertical profiles of clouds every 1.1 km along-track and the spacial resolution of each data point is 1.3 km across-track and 1.7 km along-track \cite{THECLOUDSATMISSION}. Every data point is associated with the geolocation and time of the measurement.}
One of the most interesting things about this mission is a product which provides the type of cloud among and other variables for each vertical profile \cite{cloudsat1}. 
The data acquired by CPR is stored in the CloudSat Data Processing Center, which is a SFTP server \cite{sftp_cloudsat}, in HDF-EOS format\cite{hdf}.   

\begin{figure}
\centering
\begin{subfigure}{.5\textwidth}
  \centering
  \includegraphics[width=.9\linewidth, height=7cm]{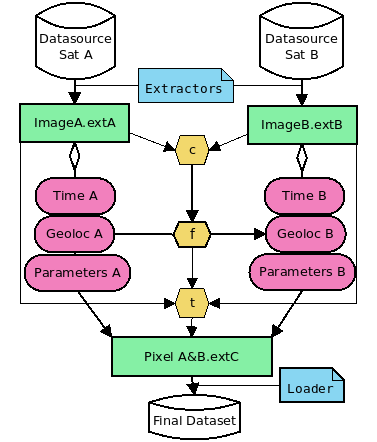}
  \caption{Conceptual design.}
  \label{fig:sub1}
\end{subfigure}%
\begin{subfigure}{.5\textwidth}
  \centering
  \includegraphics[width=.9\linewidth, height=7cm]{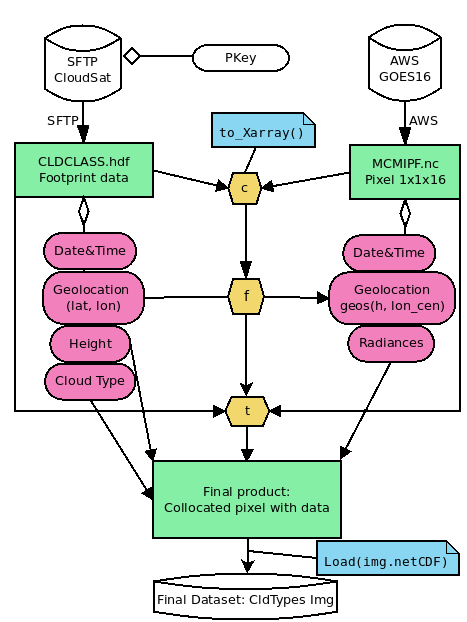}
  \caption{Stratopy workflow.}
  \label{fig:sub2}
\end{subfigure}
\caption{\jbc{Structure diagrams of the proposed pipeline built with Stratopy. On the left side, there is a diagram containing the conceptual parts, and on the right side an implementation with the components provided by the project.} }
\label{fig:diagramas}
\end{figure}

With the aforementioned descriptions, and based on the theoretical model and nomenclatures presented in the previous section, we have constructed a flowchart (Figure~\ref{fig:diagramas}) that specifies what the code does.

Figure~\ref{fig:sub2} tells the same story as Figure~\ref{fig:sub1} but in different terms, regarding the implementation. \jbc{The idea of the pipeline consists in extracting by means of a SFTP service from database A an image of an orbital passage of the CloudSat satellite (\textit{concept}), which contains in each pixel the \textit{attributes} date, time, geolocation (latitude, longitude), height in meters and type of cloud for each height. 

Once the user has selected a time range in the CloudSat extracted data, this range information is fed to the AWS extractor which retrieves a suitable multiband image from GOES16. This image is also a \textit{Concept} and in this case, each pixel has the \textit{attributes} date, time, geolocation in geostationary projection (central longitude and height of the satellite) and radiance. 
}

To know which CPR pixels or data points correspond with which ABI pixels, the Transformer $f$ makes the coordinate change: 

$$
    f: (lat, lon) \rightarrow geos(h, lon_c, R_e, R_p)
$$

where $geos$ is the projection used by GOES16 to georeference each pixel (see \cite{goeslevel1}, section 5.1.2.2 for more information). This transformation depends on GOES16 height $h$, central longitude $lon_c$, equatorial radius $R_e$ and polar radius $R_p$.
This information is used to collocate pixels from the multi-band image with pixels from the CloudSat pass and then retrieve a product where every collocated pixel contains information on the radiance, the height of the atmosphere and the cloud types found in them.

Finally, the module Loader would take care of loading this final product into storage. This module is not defined yet, but we plan to implement it as an extension of some workflow orchestration program such as Apache Airflow \cite{harenslak2021data} or Dagster \cite{ooi2005dagster};
All these technologies are based on the creation of tasks on an acyclic-directed graph (DAG) that allows fragments of the pipeline to be executed automatically in parallel or sequentially as required. 

\jbc{For your convenience, a prototype of this pipeline can be found in the Stratopy package \cite{stratopy}. Please note that the project is under active development and the current state can be explored in the project repository \url{https://github.com/paula-rj/StratoPy/tree/dev}.
}


\section{Conclusion and further work}\label{sec: conclusion}
\jbc{
The ETL formalism was very useful to achieve an orderly design of workflows. 

}
The modular structure of the concepts and attributes allows extending extractors and transformers in a simple way to extract and transform data from any of the currently active and most used Earth Observation satellites, such as GOES16/17/18 \cite{galica2016goes}, Himawari \cite{himawari}, etc.  

Also, the existence of some large ecosystem of remote sensing data analysis and data analysis packages in Python in general, and SatPy\cite{sat_data_asBD} for manipulation and transformation of data from remote-sensing earth-observing satellite instruments in particular, and Stratopy is a missing piece for orchestrating these transformations and analysis.

The future work, part of it is already started and consists of the orchestration of the processes already implemented in some kind of DAG framework such as Apache Airflow or Dagster.

%

%
%
\bibliographystyle{splncs04}
\bibliography{biblio}
%

\end{document}